\documentclass[twocolumn,showpacs,aps,prb]{revtex4-1}
\usepackage{amsmath,amssymb,graphicx,hyperref}

\newcommand{\beq}{\begin{eqnarray}}
\newcommand{\eeq}{\end{eqnarray}}
\newcommand{\beqq}{\begin{eqnarray*}}
\newcommand{\eeqq}{\end{eqnarray*}}

\usepackage{graphicx}
\usepackage{dcolumn}
\usepackage{bm}
\usepackage{color}
\usepackage{amsmath,amssymb,amsthm}
\usepackage{hyperref}

\begin{document}

\title{Self-Learning Monte Carlo Method}

\author{Junwei Liu$^{1\dag{}\ast}$, Yang Qi$^{1\dag{}}$, Zi Yang Meng$^{2}$ and Liang Fu$^{1}$}
\email{liujw@mit.edu; liangfu@mit.edu\\$^{\dagger}$The first two authors contributed equally to this work.}
\affiliation{$^1$Department of physics, Massachusetts Institute of Technology, Cambridge, MA 02139, USA}
\affiliation{$^2$Institute of Physics, Chinese Academy of Sciences, Beijing 100190, China}

\date{\today}

\begin{abstract}
Monte Carlo simulation is an unbiased numerical tool for studying classical and quantum many-body systems. One of its bottlenecks is the lack of general and efficient update algorithm for large size systems close to phase transition,
for which local updates perform badly. In this work, we propose a new general-purpose Monte Carlo method, dubbed self-learning Monte Carlo (SLMC), in which an efficient update algorithm is first learned from the training data generated in trial simulations and then used to speed up the actual simulation.
We demonstrate the efficiency of SLMC in a spin model at the phase transition point, achieving a 10-20 times speedup.
\end{abstract}


\maketitle

Monte Carlo (MC) method is a powerful and unbiased numerical tool for simulating statistical and condensed matter systems ~\cite{Binder1995,NewmanBarkema1999,Gubernatis2003,LandauBinder2005,Sandvik2010}. MC simulation obtains statistically exact values of physical observables by sampling a large number of configurations according to the Boltzmann distribution. Configurations can be generated sequentially by local update method ~\cite{Metropolis1953,HASTINGS1970}. However, when the system is close to a phase transition, 
local update can be highly inefficient as sequentially generated configurations are strongly correlated,
causing a significant slowing down in the simulation dynamics. For certain classes of models, this slowing down can be overcome by global update methods \cite{SwendsenWang1987,Wolff1989,Prokofev1998,Evertz1993,Evertz2003,Syljuaasen2002,Alet2005}, where an extensive number of local variables are changed in a single update.
However, for any generic model, it is highly challenging  to design an efficient global update method.

Inspired by great developments in machine learning~\cite{ESLBook}, in this work we propose a new approach to speed up the MC simulation. The MC sampling process generates a sequence of configurations in a Markov chain, which constitutes a massive set of data containing valuable information about the system. Meanwhile, machine learning is a powerful technique to uncover unknown properties in the data and make new predictions. Thus we expect that machine learning can extract the information hidden in the Markov chain, which we then use to improve the performance of MC simulation.

Specifically, we propose a new MC update method applicable to generic statistical models, dubbed self-learning Monte Carlo (SLMC) method. The essence of SLMC is to first perform a trial simulation with local update to obtain a sequence of configurations and their weights, serving as training data, and then to learn a rule that guides configuration update in actual simulation.
To demonstrate the power of SLMC, we study a statistical model (see Eq.~\ref{eq:H}), for which no efficient global update scheme is known.
We find that in comparison to the local update, SLMC significantly reduces the autocorrelation time, especially near the phase transition.

{\it Outline of SLMC~~} Before presenting our method, let us recall that configurations in MC simulation can be updated through a Markov process, where the transition probability from configuration $A$ to $B$, $P(A\rightarrow B)$, is required to satisfy the detailed balance principle (DBP)\cite{Metropolis1953}, $P(A\rightarrow B)/P(B\rightarrow A)= W(B)/W(A)$, where $W$ is the probability distribution of configurations. Update methods can be roughly divided into two types: local and global.

Local update is a general-purpose, model-independent method, consisting of two steps. First, one randomly chooses a single site in the current configuration and proposes a new configuration by changing the variable on this site. Second, one decides whether the proposed move is accepted or rejected based on DBP. If accepted, the next configuration in Markov chain will be the new one; otherwise it will be a copy of the current one.
Clearly, the way a local move is proposed in the first step is completely general and does not use any knowledge of the model.
Local update works well for many systems, but suffers heavily from the critical slowing down close to phase transitions~\cite{SwendsenWang1987,Wolff1989}.
In such cases, the autocorrelation time within the Markov chain $\tau$ becomes very large, and in fact diverges with the system size $L$ as $\tau \sim \tau_0 L^z$ at critical points, where $z$ is the dynamical exponent of MC simulation.

To overcome the dramatic increase of autocorrelation time for local update, many global update methods have been developed, such as Swendsen-Wang~\cite{SwendsenWang1987}, Wolff~\cite{Wolff1989}, worm~\cite{Prokofev1998}, loop~\cite{Evertz1993,Evertz2003} and directed loop~\cite{Syljuaasen2002,Alet2005} algorithms. In all these methods, variables on an extensive number of sites are simultaneously changed in a single MC update, thus reducing the dynamic exponent $z$ significantly. However, unlike the local update, here the proposal of a trial configuration and the determination of its acceptance are intricately linked, because the proposed move already takes into account the DBP. Thus global updates are ingeniously designed methods targeted for special models. For a given generic model, it is very difficult to design an efficient global update method.

\begin{figure}[tbp]
\includegraphics[width=\columnwidth]{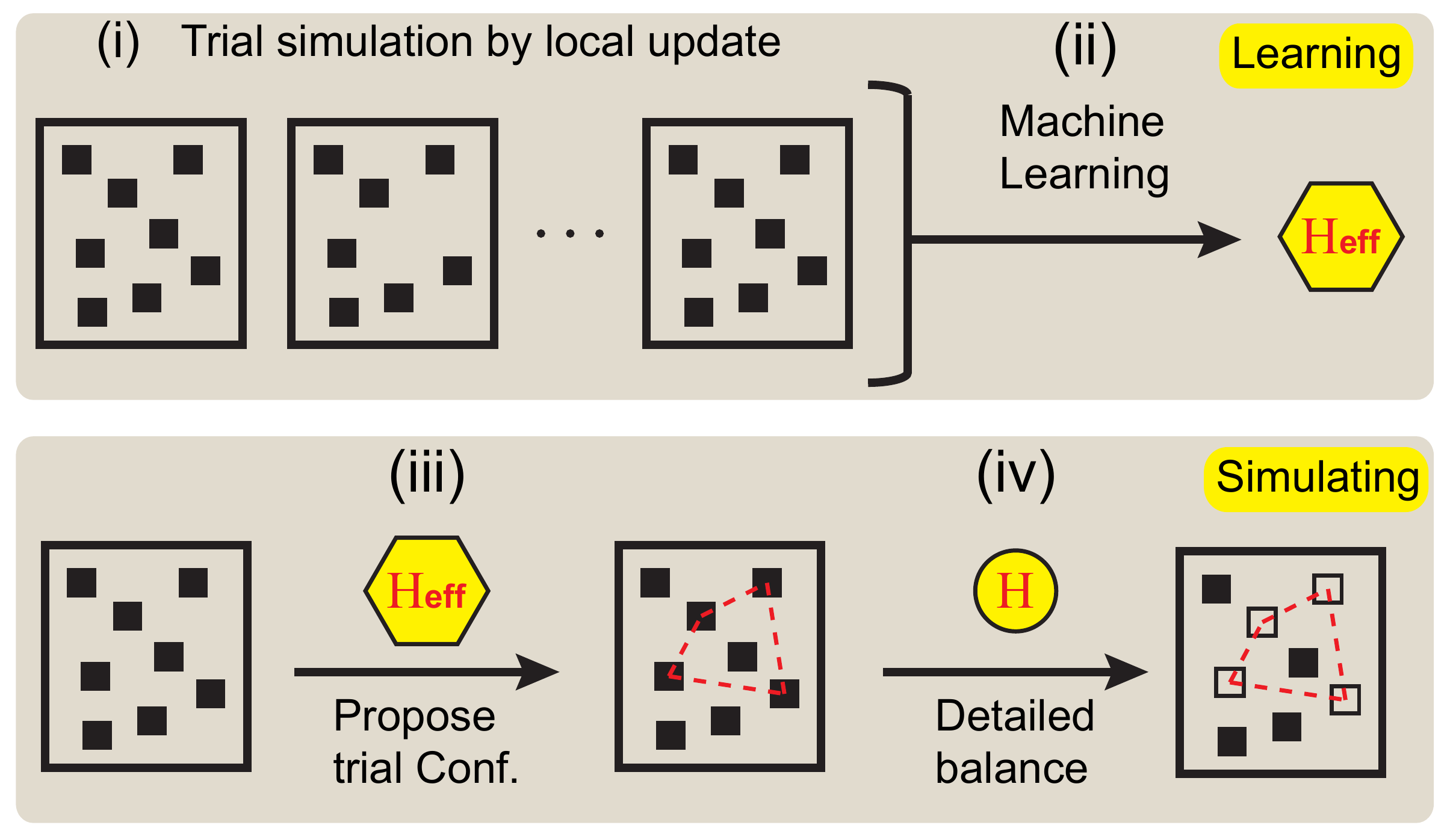}
\caption{(color online) Schematic illustration of learning process (top panel) and simulating process (bottom panel) in self-learning Monte Carlo.}
\label{fig:schematic}
\end{figure}

From the comparison of local and global updates, we conclude that a general-purpose MC update method that can outperform local update must satisfy the following requirements: (1) a large number of sites should be involved in each move that updates the current configuration; (2) the proposal and the acceptance of moves should be independent. For systems at the critical point, we further require the number of sites involved in each move increase with the system size in order to reduce the dynamical exponent $z$ in MC simulation.

Guided by these requirements, we now propose the detailed procedure of SLMC method. As shown in Fig.~\ref{fig:schematic},  SLMC consists of four steps: (i) perform a trial MC simulation using local update to generate a large number of  configurations, which serve as training data; (ii) learn an effective Hamiltonian $H_{\text{eff}}$ from this training data; (iii) propose moves according to $H_{\text{eff}}$ in the  actual MC simulation; (iv) determine whether the proposed moves will be accepted or rejected based on the detailed balance principle of the {\it original} Hamiltonian $H$. Steps (i) and (ii) constitute the learning process, whereas steps (iii) and (iv) are repeated in the actual MC simulation to calculate physical observables.

We further outline how to implement step (ii) and (iii) in actual simulations for a model to be presented below. We use machine learning~\cite{ESLBook} in step (ii) to train an effective Hamiltonian, which can be efficiently simulated using a global update method even though the original Hamiltonian cannot. Then step (iii) can be easily implemented using this global update.

{\it Model and results:}
To demonstrate the power of SLMC, we study a classical model on a 2D square lattice
\beq\label{eq:H}
H= - J \sum_{\langle ij\rangle} S_i S_j - K \sum_{ijkl \in \qedsymbol } S_i S_j S_k S_l,
\eeq
where $S_i=\pm1$ is the Ising spin on site $i$. $J$ is the nearest neighbor (NN) interaction and $K$ is the interaction among the four spins in the same plaquette. We set ferromagnetic interactions, i.e., $J>0$ and $K>0$. For any finite $J$ and $K$, there is a phase transition from paramagnetic phase at high temperature to ferromagnetic phase at low temperature, which belongs to the 2D Ising universality class. For $K=0$, this model reduces to the standard Ising model which can be simulated efficiently by the Wolff method. However, for $K \neq 0$, no simple and efficient global update method is known. Below we will show that SLMC method significantly reduces the autocorrelation time near the critical point, using $K/J=0.2$ as an example. More results can be found in the Supplemental Material (SM) \cite{sm}.

\begin{table}[]
\centering
\caption{The trained parameters $\{\tilde{J}_n\}$ of the effective model in Eq.~\ref{eq:Heff}, without and with setting $\tilde{J}_n = 0$ ($n \geq 2$).  }
\label{Jeff}
\begin{tabular*}{\columnwidth}{@{\extracolsep{\fill}}ccccc}
  \hline\hline
          & $\tilde{J}_1$ & $\tilde{J}_2$  & $\tilde{J}_3$  & Mean error \\ \hline
  Train 1 & 1.2444        & -0.0873        & -0.0120        & 0.0009\\
  Train 2 & 1.1064        & -              & -              & 0.0011\\
  \hline\hline
\end{tabular*}
\end{table}

As outlined before, the initial step of the SLMC is to train an effective Hamiltonian, $H_{\text{eff}}$, from a sample of configurations generated by local update based on the original Hamiltonian in Eq.~\ref{eq:H}. We choose $H_{\text{eff}}$ to be a generalized Ising Hamiltonian with two-body spin interactions over various ranges,
\begin{equation}
  \label{eq:Heff}
  H_{\text{eff}} = E_0 - \tilde{J}_1\sum_{\langle ij \rangle _1} S_i S_j- \tilde{J}_2\sum_{\langle ij \rangle_2} S_i S_j - \dots ,
\end{equation}
where $\langle ij \rangle_n$ denotes the $n$-th NN interaction and $\tilde{J}_{n}$ is the corresponding interaction parameter.

We now train $H_{\text{eff}}$ from the training sample by optimizing $E_0$ and $\{\tilde{J}_{n}\}$. In principle, this can be viewed as an unsupervised learning process~\cite{ESLBook,torlai2016}, where a new statistical model $H_{\text{eff}}$ is trained using a subset of features extracted from the configurations. However, by taking advantage of knowing $H$ for each configuration, we can more efficiently train $H_{\text{eff}}$ through a simple and reliable linear regression. For the $a$-th configuration in the sample, we compute its energy $E^a$ [from Eq.~\ref{eq:H}] and all the $n$-th NN spin-spin correlations $C_n^a=\sum_{\langle ij\rangle_n}S_iS_j$, which serve as the actual training data. Then, $E_0$ and $\{\tilde{J}_n\}$ can be easily trained from a multi-linear regression of $E^a$ and $\{C_n^a\}$, $E^a = \sum_n \tilde{J}_n C_n^a + E_0$.
The results are as shown in Table~\ref{Jeff} (Train 1). It is clear that $\tilde{J}_1$ is dominant and much larger than others, which implies we could set $\tilde{J}_n = 0$ ($n \geq 2$). And then, by a linear regression, we can successfully extract the most optimized $\tilde{J}_1$ (Train 2 in Table~\ref{Jeff}). It is found that the mean error is almost the same to the case without setting $\tilde{J}_n = 0$ ($n \geq 2$), which is expected since all $\tilde{J}_n$ ($n \geq 2$) obtained from the multi-linear regression are negligible. Through this training process, we conclude that only the nearest interaction is relevant there, thus we only keep this term in the following simulations. We emphasize that this trained model $H_{\text{eff}}$ only approximates the original one for the configurations that are statistically significant in the sample, i.e., the ones near the free energy minimum. Thus $H_{\text{eff}}$ can be regarded as an effective model.
We notice that, recently, there are many other attempts to apply machine learning to MC simulations~\cite{carrasquilla2016,WangLei2016,Carleo2016,Broecker2016,Chng2016,changlani2015}.

In addition, it should be addressed that the training of $H_{\text{eff}}$ could be self-improved by a reinforced learning process. Usually, a good initial sample could be very hard to generate using only local update, especially for systems at the critical temperature $T_c$ or with strong fluctuation. In this case, we first train an effective model $H_{\text{eff}}$ using a simulation at temperature $T>T_c$, and then generate another sample at $T_c$, using the self-learning update with $H_{\text{eff}}$ learned from the first iteration. Later, a more accurate $H_{\text{eff}}$ can be learned from the second-iteration sample. In actual simulations, one can further improve this process by using more iterations, each done with a smaller sample. More details can be found in the Supplemental Material.

\begin{figure}
  \includegraphics[width=\columnwidth]{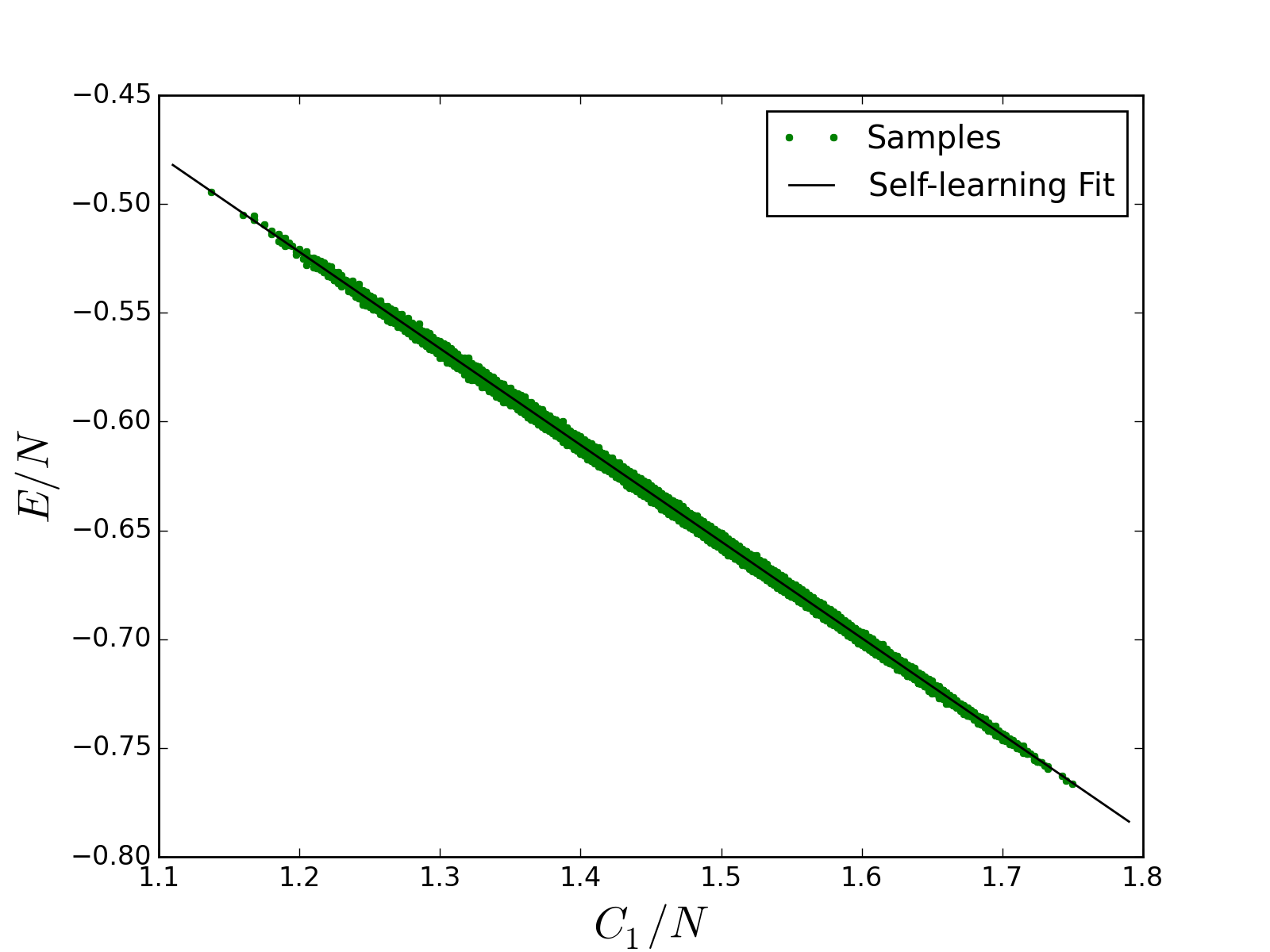}
  \caption{\label{fig:dist_fit}(color online) Fitting of the distribution drawn from a sample of configurations in a Markov chain. The green dots represent configurations in the sample, for which the $x$ axis shows the feature of the nearest-neighbor spin-spin correlation $C_1$, and the $y$ axis shows the energy (per site) $E/N$ computed from the original model in Eq.~\ref{eq:H}.
  }
\end{figure}

Through this iterative training process, we successfully arrive at the final $H_{\text{eff}}$. As shown in Fig. 2, $H_{\text{eff}}$ (Self-Learning Fit) indeed fits the energy of the configurations that are statistically significant in the simulation. In the main part of the figure, the data points are concentrated in the vicinity of the fitted line, indicating that trained $H_{\text{eff}}$ is indeed a good description of the low-energy physics.

Following the procedure of SLMC, once training process is finished, cluster update with the Wolff algorithm according to $H_{\text{eff}}$ can be constructed. Then, the generated cluster update is accepted or rejected with a probability accounting for the energy difference between the effective model and the original model. The probability of accepting a cluster is as follows,
\begin{equation}
  \label{eq:paccept}
  \alpha(A\rightarrow B) = \min\{1,
   e^{-\beta[(E_B-E_B^{\text{eff}})-(E_A-E_A^{\text{eff}})]}\},
\end{equation}
where $A$ and $B$ denote the configurations before and after flipping the cluster. $E_A$ and $E_A^{\text{eff}}$ denote the energies of a configuration $A$, for the original model in Eq.~\ref{eq:H} and the effective model in Eq.~\ref{eq:Heff}, respectively. Derivation of Eq.~\ref{eq:paccept} can be found in the SM\cite{sm}. With Eq.~\ref{eq:paccept}, the detailed balance is satisfied, and the SLMC is \emph{exact}, despite the use of an approximate effective model in constructing the cluster.

\begin{figure}
  \includegraphics[width=\columnwidth]{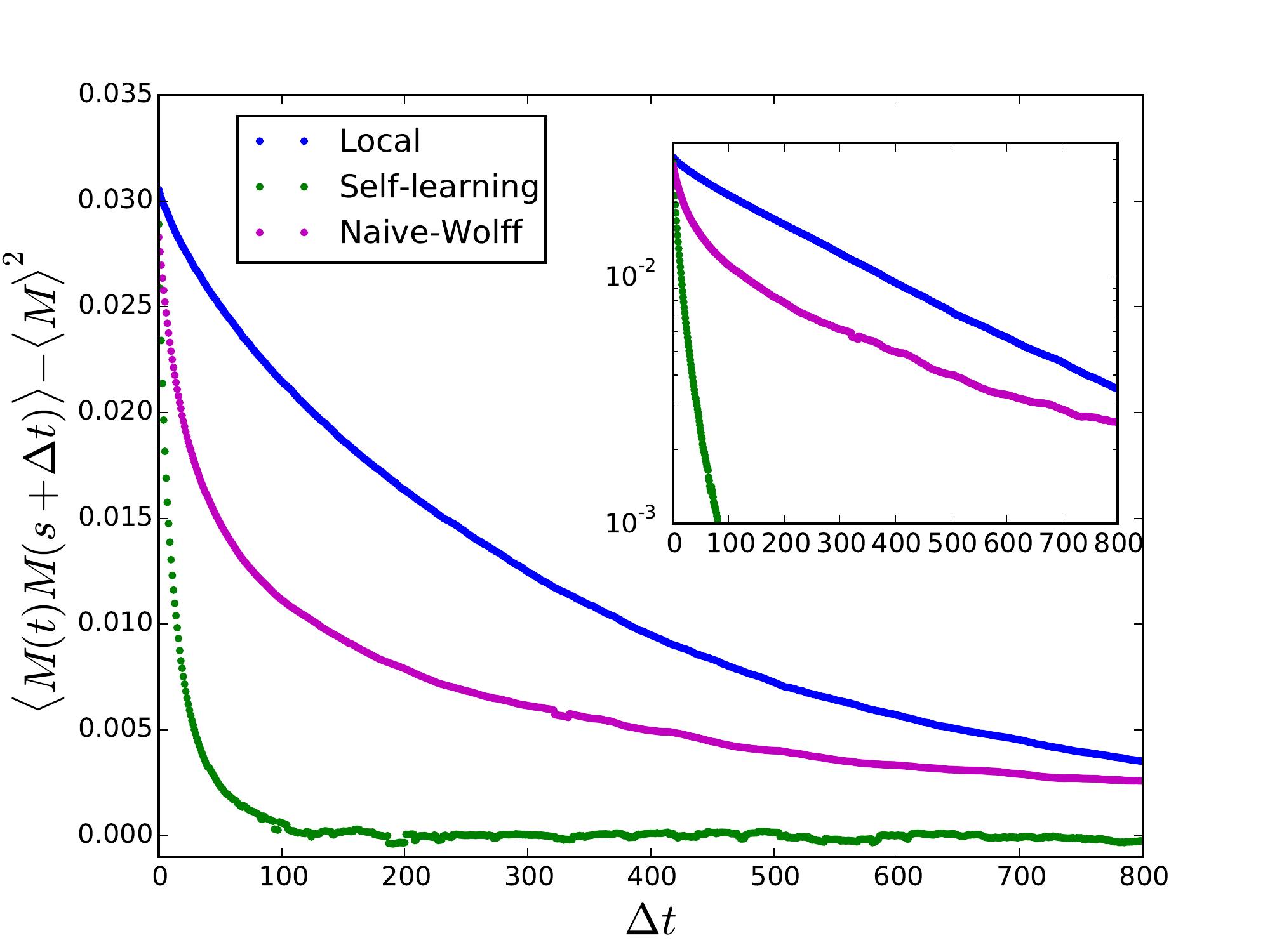}
  \caption{\label{fig:corr_compare}(color online) The decay of autocorrelation functions as a function of MC steps, obtained using different update algorithms. Inset, semi-log plot of the same data.}
\end{figure}

To test the efficiency of the update scheme in SLMC, we measure the autocorrelation time $\tau$, which signifies how correlated the MC configurations are in the Markov chain (detailed relation of $\tau$ with the computational complexity of MC algorithm can be found in SM \cite{sm}). In Fig.~\ref{fig:corr_compare}, we plot $\tau$ of the ferromagnetic order parameter $M =\frac1N\left|\sum_iS_i\right|$, where $N$ is the number of sites, measured at each step of Markov chain, generated by different update algorithms on a square lattice of linear size $L=40$. The simulation is done at $T_c$, which is determined by the Binder ratio as shown in SM \cite{sm}.

We compare results of the local update, the self-learning update using $H_{\text{eff}}$ and also a naive Wolff-like cluster update with the bare two-body $J$ term from the original model in Eq.~\ref{eq:H} is used to construct a cluster. The autocorrelation functions generated by all updates decay with the MC steps $\Delta t$, and autocorrelation time $\tau$ can be obtained from fitting in the form of $e^{-\Delta t/\tau}$. Our results show that comparing to the local and naive cluster updates, the self-learning update has the much shorter $\tau$. In particular, at this system size, the self-learning update is about 24-times faster than the local update, while the naive Wolff-like cluster update does not gain much speed-up.

\begin{figure}
  \includegraphics[width=\columnwidth]{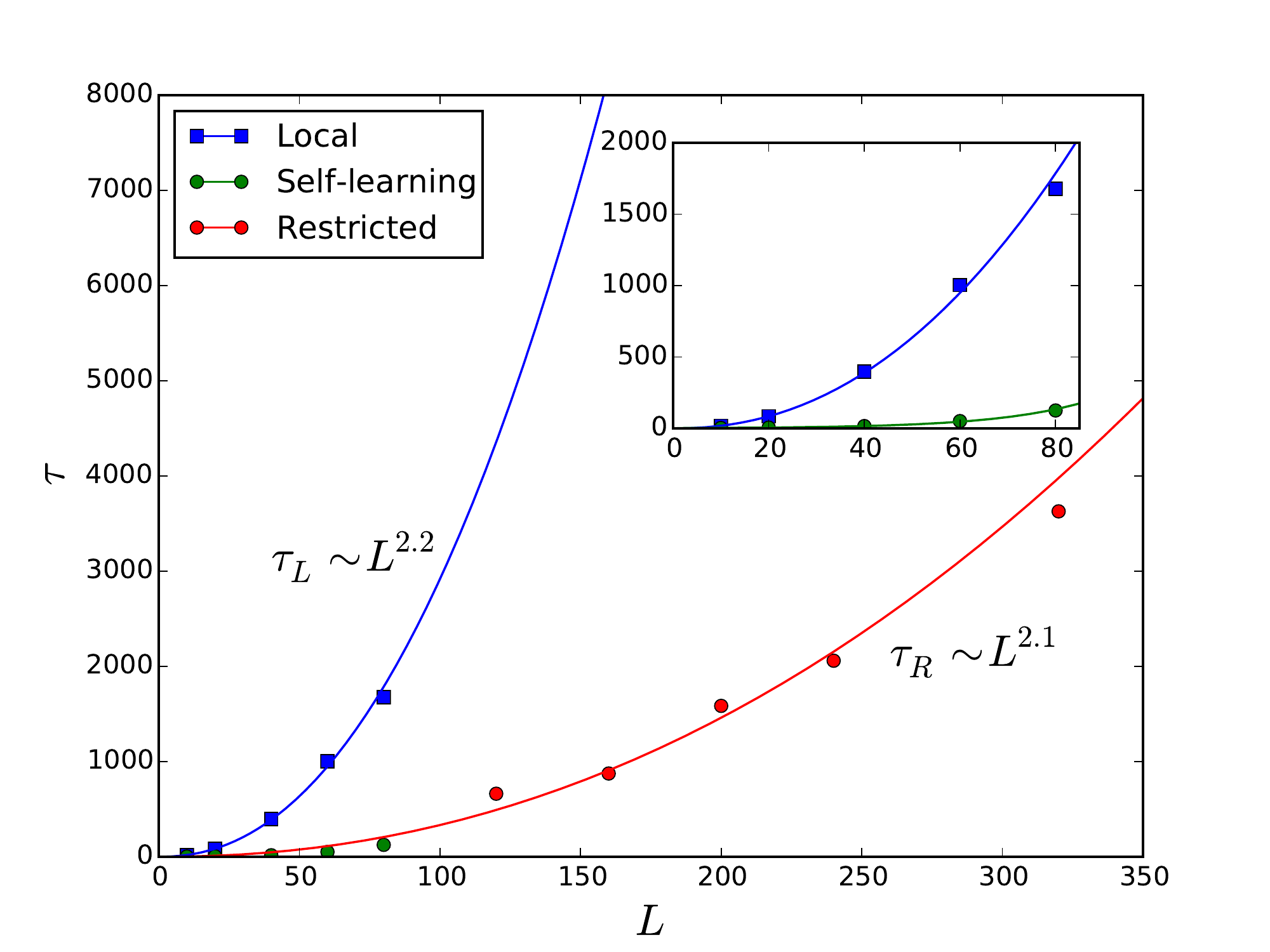}
  \caption{\label{fig:auto_compare}(color online) The scaling behavior of autocorrelation times of local update $\tau_L$, SLMC update $\tau_S$, and the restricted SLMC update $\tau_R$. Inset is a zooming for $L<80$.}
\end{figure}

While Fig.~\ref{fig:corr_compare} is an example of the better performance of SLMC for a fixed system size at $T_c$, we have further collected the autocorrelation time $\tau$ at $T_c$ for local and self-learning updates with many different system sizes, and hence extract the scaling behavior of $\tau$ with respect to $L$. The results are shown in Fig.~\ref{fig:auto_compare}. The blue squares are the $\tau_{L}$, i.e., autocorrelation time for local update, and it follows $\tau_{L}\sim L^{2.2}$, well consistent with literature on critical slowing down~\cite{SwendsenWang1987,Wolff1989}. The green dots are the $\tau_{S}$, i.e., autocorrelation time for self-learning update. For all the tested systems size $L \leq 80$, the $\tau_{S}$ delivers a large speedup about 20 times (see inset of Fig.~\ref{fig:auto_compare} for clarity).

For very large system size, we find $\tau_{S}$ increases exponentially with $L$, $\tau_S\propto e^{L/L_0}$ (more details in SM \cite{sm}). This is because of a finite energy difference between the effective model in Eq.~\ref{eq:Heff} and the original model in Eq.~\ref{eq:H}. Therefore, the acceptance ratio of flipping the whole cluster in Eq.~\ref{eq:paccept} decreases exponentially as the length of cluster boundary grows with increasing $L$, which renders the exponential increase of the autocorrelation time. But this drawback in SLMC can be easily remedied by simply restricting the maximum size of the cluster in Wolff algorithm~\cite{Barkema1993}. With this improvement, the averaged acceptance ratio can be expected to be fixed and SLMC should have the same scaling function for autocorrelation time as local update, $\tau_R=\tau_0 L^z$. However, by tuning the maximum size of cluster, we can achieve a much smaller prefactor $\tau_0$, and the optimized maximum cluster size can be automatically self-learned via a model-independent procedure (more details in SM\cite{sm}). This is indeed the case. As shown by the red dots in Fig.~\ref{fig:auto_compare},  when the growth of the cluster is restricted to an area within 40 lattice spacing, the autocorrelation time $\tau_R$ becomes $\tau_R\propto L^{2.1}$, which obeys the same power law as $\tau_L$, but with a prefactor about 10 times smaller (More details about the design of this restricted SLMC is provided in SM \cite{sm}). Therefore, although SLMC still suffers from the critical slowing down in the thermodynamic limit, we can gain a 10-fold speedup. That means SLMC can achieve much larger system size than local update, which helps to overcome the finite size effect. Moreover, for medium-size systems, the SLMC without restriction can easily gain a 20-fold speedup, as shown by $\tau_{S}$.

{\it Discussion:} We now discuss the applicability of SLMC method to a broader class of problems in statistical and condensed matter systems. Besides spin systems, many models of great interest may be transformed into spin models with short-range interactions \cite{AssaadEvertz2008,Sandvik2010}, for which efficient global update methods are available. In such cases, SLMC can be readily implemented similar to our model studied above. In particular, we expect SLMC to be very useful for studying strongly correlated fermion systems~\cite{liu2016fermion,xu2016}, where no efficient global update method is currently known. Moreover, by employing rapidly-developing machine learning techniques, SLMC method may be able to learn configuration update on its own, without relying on a given effective Hamiltonian. If realized, this will further increase the efficiency and versatility of SLMC.

SLMC may also bridge numerical and theoretical studies. The effective Hamiltonian trained or learned from the MC simulation may guide the theoretical study of the original model. The benefit is mutual: theoretical understanding may improve the accuracy of the effective model and thus the performance of numerical simulation.

{\it Note added}: Recently we noted a related work \cite{huang2016accelerate}.

{\it Acknowledgement:}
We thank Lanjun Wang and Xiao-Yan Xu for helpful discussions. This work is supported by the DOE Office of Basic Energy Sciences, Division of Materials Sciences and Engineering under Award DE-SC0010526. LF is supported partly by the David and Lucile Packard Foundation. ZYM is supported by the Ministry of Science and Technology (MOST) of China under Grant No. 2016YFA0300502, the National Natural Science Foundation of China (NSFC Grants No. 11421092 and No. 11574359), and the National Thousand-Young-Talents Program of China. JWL and QY thank the hospitality of Institute for Physics, Chinese Academy of Sciences where part of the work is performed.

\bibliography{MLMC_Ref}

\begin{thebibliography}{28}%
\makeatletter
\providecommand \@ifxundefined [1]{%
 \@ifx{#1\undefined}
}%
\providecommand \@ifnum [1]{%
 \ifnum #1\expandafter \@firstoftwo
 \else \expandafter \@secondoftwo
 \fi
}%
\providecommand \@ifx [1]{%
 \ifx #1\expandafter \@firstoftwo
 \else \expandafter \@secondoftwo
 \fi
}%
\providecommand \natexlab [1]{#1}%
\providecommand \enquote  [1]{``#1''}%
\providecommand \bibnamefont  [1]{#1}%
\providecommand \bibfnamefont [1]{#1}%
\providecommand \citenamefont [1]{#1}%
\providecommand \href@noop [0]{\@secondoftwo}%
\providecommand \href [0]{\begingroup \@sanitize@url \@href}%
\providecommand \@href[1]{\@@startlink{#1}\@@href}%
\providecommand \@@href[1]{\endgroup#1\@@endlink}%
\providecommand \@sanitize@url [0]{\catcode `\\12\catcode `\$12\catcode
  `\&12\catcode `\#12\catcode `\^12\catcode `\_12\catcode `\%12\relax}%
\providecommand \@@startlink[1]{}%
\providecommand \@@endlink[0]{}%
\providecommand \url  [0]{\begingroup\@sanitize@url \@url }%
\providecommand \@url [1]{\endgroup\@href {#1}{\urlprefix }}%
\providecommand \urlprefix  [0]{URL }%
\providecommand \Eprint [0]{\href }%
\providecommand \doibase [0]{http://dx.doi.org/}%
\providecommand \selectlanguage [0]{\@gobble}%
\providecommand \bibinfo  [0]{\@secondoftwo}%
\providecommand \bibfield  [0]{\@secondoftwo}%
\providecommand \translation [1]{[#1]}%
\providecommand \BibitemOpen [0]{}%
\providecommand \bibitemStop [0]{}%
\providecommand \bibitemNoStop [0]{.\EOS\space}%
\providecommand \EOS [0]{\spacefactor3000\relax}%
\providecommand \BibitemShut  [1]{\csname bibitem#1\endcsname}%
\let\auto@bib@innerbib\@empty
\bibitem [{\citenamefont {Binder}(1995)}]{Binder1995}%
  \BibitemOpen
  \bibinfo {editor} {\bibfnamefont {K.}~\bibnamefont {Binder}},\ ed.,\ \href
  {\doibase 10.1007/3-540-60174-0} {\emph {\bibinfo {title} {The Monte Carlo
  Method in Condensed Matter Physics}}}\ (\bibinfo  {publisher}
  {Springer-Verlag Berlin Heidelberg},\ \bibinfo {year} {1995})\BibitemShut
  {NoStop}%
\bibitem [{\citenamefont {Newman}\ and\ \citenamefont
  {Barkema}(1999)}]{NewmanBarkema1999}%
  \BibitemOpen
  \bibfield  {author} {\bibinfo {author} {\bibfnamefont {M.~E.~J.}\
  \bibnamefont {Newman}}\ and\ \bibinfo {author} {\bibfnamefont {G.~T.}\
  \bibnamefont {Barkema}},\ }\href
  {https://global.oup.com/academic/product/monte-carlo-methods-in-statistical-physics-9780198517979?cc=cn&lang=en&#}
  {\emph {\bibinfo {title} {Monte Carlo Methods in Statistical Physics}}}\
  (\bibinfo  {publisher} {Oxford University Press},\ \bibinfo {year}
  {1999})\BibitemShut {NoStop}%
\bibitem [{\citenamefont {Gubernatis}(2003)}]{Gubernatis2003}%
  \BibitemOpen
  \bibinfo {editor} {\bibfnamefont {J.~E.}\ \bibnamefont {Gubernatis}},\ ed.,\
  \href {http://scitation.aip.org/content/aip/proceeding/aipcp/690} {\emph
  {\bibinfo {title} {THE MONTE CARLO METHOD IN THE PHYSICAL SCIENCES:
  Celebrating the 50th Anniversary of the Metropolis Algorithm}}},\ Vol.\
  \bibinfo {volume} {690},\ \bibinfo {organization} {AIP Conference
  Proceedings}\ (\bibinfo  {publisher} {AIP Publishing},\ \bibinfo {address}
  {Los Alamos, New Mexico (USA)},\ \bibinfo {year} {2003})\BibitemShut
  {NoStop}%
\bibitem [{\citenamefont {Landau}\ and\ \citenamefont
  {Binder}(2005)}]{LandauBinder2005}%
  \BibitemOpen
  \bibfield  {author} {\bibinfo {author} {\bibfnamefont {D.}~\bibnamefont
  {Landau}}\ and\ \bibinfo {author} {\bibfnamefont {K.}~\bibnamefont
  {Binder}},\ }\href {https://books.google.ca/books?id=HbxQxS7tHiYC} {\emph
  {\bibinfo {title} {A Guide to Monte Carlo Simulations in Statistical
  Physics}}}\ (\bibinfo  {publisher} {Cambridge University Press},\ \bibinfo
  {year} {2005})\BibitemShut {NoStop}%
\bibitem [{\citenamefont {Sandvik}(2010)}]{Sandvik2010}%
  \BibitemOpen
  \bibfield  {author} {\bibinfo {author} {\bibfnamefont {A.~W.}\ \bibnamefont
  {Sandvik}},\ }\href {\doibase http://dx.doi.org/10.1063/1.3518900} {\bibfield
   {journal} {\bibinfo  {journal} {AIP Conference Proceedings}\ }\textbf
  {\bibinfo {volume} {1297}},\ \bibinfo {pages} {135} (\bibinfo {year}
  {2010})}\BibitemShut {NoStop}%
\bibitem [{\citenamefont {Metropolis}\ \emph {et~al.}(1953)\citenamefont
  {Metropolis}, \citenamefont {Rosenbluth}, \citenamefont {Rosenbluth},
  \citenamefont {Teller},\ and\ \citenamefont {Teller}}]{Metropolis1953}%
  \BibitemOpen
  \bibfield  {author} {\bibinfo {author} {\bibfnamefont {N.}~\bibnamefont
  {Metropolis}}, \bibinfo {author} {\bibfnamefont {A.~W.}\ \bibnamefont
  {Rosenbluth}}, \bibinfo {author} {\bibfnamefont {M.~N.}\ \bibnamefont
  {Rosenbluth}}, \bibinfo {author} {\bibfnamefont {A.~H.}\ \bibnamefont
  {Teller}}, \ and\ \bibinfo {author} {\bibfnamefont {E.}~\bibnamefont
  {Teller}},\ }\href {\doibase http://dx.doi.org/10.1063/1.1699114} {\bibfield
  {journal} {\bibinfo  {journal} {J. Chem. Phys.}\ }\textbf {\bibinfo {volume}
  {21}},\ \bibinfo {pages} {1087} (\bibinfo {year} {1953})}\BibitemShut
  {NoStop}%
\bibitem [{\citenamefont {Hastings}(1970)}]{HASTINGS1970}%
  \BibitemOpen
  \bibfield  {author} {\bibinfo {author} {\bibfnamefont {W.~K.}\ \bibnamefont
  {Hastings}},\ }\href {\doibase 10.1093/biomet/57.1.97} {\bibfield  {journal}
  {\bibinfo  {journal} {Biometrika}\ }\textbf {\bibinfo {volume} {57}},\
  \bibinfo {pages} {97} (\bibinfo {year} {1970})}\BibitemShut {NoStop}%
\bibitem [{\citenamefont {Swendsen}\ and\ \citenamefont
  {Wang}(1987)}]{SwendsenWang1987}%
  \BibitemOpen
  \bibfield  {author} {\bibinfo {author} {\bibfnamefont {R.~H.}\ \bibnamefont
  {Swendsen}}\ and\ \bibinfo {author} {\bibfnamefont {J.-S.}\ \bibnamefont
  {Wang}},\ }\href {\doibase 10.1103/PhysRevLett.58.86} {\bibfield  {journal}
  {\bibinfo  {journal} {Phys. Rev. Lett.}\ }\textbf {\bibinfo {volume} {58}},\
  \bibinfo {pages} {86} (\bibinfo {year} {1987})}\BibitemShut {NoStop}%
\bibitem [{\citenamefont {Wolff}(1989)}]{Wolff1989}%
  \BibitemOpen
  \bibfield  {author} {\bibinfo {author} {\bibfnamefont {U.}~\bibnamefont
  {Wolff}},\ }\href {\doibase 10.1103/PhysRevLett.62.361} {\bibfield  {journal}
  {\bibinfo  {journal} {Phys. Rev. Lett.}\ }\textbf {\bibinfo {volume} {62}},\
  \bibinfo {pages} {361} (\bibinfo {year} {1989})}\BibitemShut {NoStop}%
\bibitem [{\citenamefont {Prokof'ev}\ \emph {et~al.}(1998)\citenamefont
  {Prokof'ev}, \citenamefont {Svistunov},\ and\ \citenamefont
  {Tupitsyn}}]{Prokofev1998}%
  \BibitemOpen
  \bibfield  {author} {\bibinfo {author} {\bibfnamefont {N.}~\bibnamefont
  {Prokof'ev}}, \bibinfo {author} {\bibfnamefont {B.}~\bibnamefont
  {Svistunov}}, \ and\ \bibinfo {author} {\bibfnamefont {I.}~\bibnamefont
  {Tupitsyn}},\ }\href {\doibase
  http://dx.doi.org/10.1016/S0375-9601(97)00957-2} {\bibfield  {journal}
  {\bibinfo  {journal} {Phys. Lett. A}\ }\textbf {\bibinfo {volume} {238}},\
  \bibinfo {pages} {253 } (\bibinfo {year} {1998})}\BibitemShut {NoStop}%
\bibitem [{\citenamefont {Evertz}\ \emph {et~al.}(1993)\citenamefont {Evertz},
  \citenamefont {Lana},\ and\ \citenamefont {Marcu}}]{Evertz1993}%
  \BibitemOpen
  \bibfield  {author} {\bibinfo {author} {\bibfnamefont {H.~G.}\ \bibnamefont
  {Evertz}}, \bibinfo {author} {\bibfnamefont {G.}~\bibnamefont {Lana}}, \ and\
  \bibinfo {author} {\bibfnamefont {M.}~\bibnamefont {Marcu}},\ }\href
  {\doibase 10.1103/PhysRevLett.70.875} {\bibfield  {journal} {\bibinfo
  {journal} {Phys. Rev. Lett.}\ }\textbf {\bibinfo {volume} {70}},\ \bibinfo
  {pages} {875} (\bibinfo {year} {1993})}\BibitemShut {NoStop}%
\bibitem [{\citenamefont {Evertz}(2003)}]{Evertz2003}%
  \BibitemOpen
  \bibfield  {author} {\bibinfo {author} {\bibfnamefont {H.~G.}\ \bibnamefont
  {Evertz}},\ }\href {\doibase 10.1080/0001873021000049195} {\bibfield
  {journal} {\bibinfo  {journal} {Advances in Physics}\ }\textbf {\bibinfo
  {volume} {52}},\ \bibinfo {pages} {1} (\bibinfo {year} {2003})}\BibitemShut
  {NoStop}%
\bibitem [{\citenamefont {Sylju\aa{}sen}\ and\ \citenamefont
  {Sandvik}(2002)}]{Syljuaasen2002}%
  \BibitemOpen
  \bibfield  {author} {\bibinfo {author} {\bibfnamefont {O.~F.}\ \bibnamefont
  {Sylju\aa{}sen}}\ and\ \bibinfo {author} {\bibfnamefont {A.~W.}\ \bibnamefont
  {Sandvik}},\ }\href {\doibase 10.1103/PhysRevE.66.046701} {\bibfield
  {journal} {\bibinfo  {journal} {Phys. Rev. E}\ }\textbf {\bibinfo {volume}
  {66}},\ \bibinfo {pages} {046701} (\bibinfo {year} {2002})}\BibitemShut
  {NoStop}%
\bibitem [{\citenamefont {Alet}\ \emph {et~al.}(2005)\citenamefont {Alet},
  \citenamefont {Wessel},\ and\ \citenamefont {Troyer}}]{Alet2005}%
  \BibitemOpen
  \bibfield  {author} {\bibinfo {author} {\bibfnamefont {F.}~\bibnamefont
  {Alet}}, \bibinfo {author} {\bibfnamefont {S.}~\bibnamefont {Wessel}}, \ and\
  \bibinfo {author} {\bibfnamefont {M.}~\bibnamefont {Troyer}},\ }\href
  {\doibase 10.1103/PhysRevE.71.036706} {\bibfield  {journal} {\bibinfo
  {journal} {Phys. Rev. E}\ }\textbf {\bibinfo {volume} {71}},\ \bibinfo
  {pages} {036706} (\bibinfo {year} {2005})}\BibitemShut {NoStop}%
\bibitem [{\citenamefont {Hastie}\ \emph {et~al.}(2009)\citenamefont {Hastie},
  \citenamefont {Tibshirani},\ and\ \citenamefont {Friedman}}]{ESLBook}%
  \BibitemOpen
  \bibfield  {author} {\bibinfo {author} {\bibfnamefont {T.}~\bibnamefont
  {Hastie}}, \bibinfo {author} {\bibfnamefont {R.}~\bibnamefont {Tibshirani}},
  \ and\ \bibinfo {author} {\bibfnamefont {J.}~\bibnamefont {Friedman}},\
  }\href {\doibase 10.1007/978-0-387-84858-7} {\emph {\bibinfo {title} {The
  Elements of Statistical Learning}}}\ (\bibinfo  {publisher} {Springer New
  York},\ \bibinfo {year} {2009})\BibitemShut {NoStop}%
\bibitem [{sm()}]{sm}%
  \BibitemOpen
  \href@noop {} {}\bibinfo {note} {See Supplemental Material at URL for more
  details and results.}\BibitemShut {Stop}%
\bibitem [{\citenamefont {Torlai}\ and\ \citenamefont
  {Melko}(2016)}]{torlai2016}%
  \BibitemOpen
  \bibfield  {author} {\bibinfo {author} {\bibfnamefont {G.}~\bibnamefont
  {Torlai}}\ and\ \bibinfo {author} {\bibfnamefont {R.~G.}\ \bibnamefont
  {Melko}},\ }\href {https://arxiv.org/abs/1606.02718} {\bibfield  {journal}
  {\bibinfo  {journal} {arXiv:1606.02718}\ } (\bibinfo {year}
  {2016})}\BibitemShut {NoStop}%
\bibitem [{\citenamefont {Carrasquilla}\ and\ \citenamefont
  {Melko}(2016)}]{carrasquilla2016}%
  \BibitemOpen
  \bibfield  {author} {\bibinfo {author} {\bibfnamefont {J.}~\bibnamefont
  {Carrasquilla}}\ and\ \bibinfo {author} {\bibfnamefont {R.~G.}\ \bibnamefont
  {Melko}},\ }\href {https://arxiv.org/abs/1605.01735} {\bibfield  {journal}
  {\bibinfo  {journal} {arXiv:1605.01735}\ } (\bibinfo {year}
  {2016})}\BibitemShut {NoStop}%
\bibitem [{\citenamefont {Wang}(2016)}]{WangLei2016}%
  \BibitemOpen
  \bibfield  {author} {\bibinfo {author} {\bibfnamefont {L.}~\bibnamefont
  {Wang}},\ }\href {https://arxiv.org/abs/1606.00318} {\bibfield  {journal}
  {\bibinfo  {journal} {arXiv:1606.00318}\ } (\bibinfo {year}
  {2016})}\BibitemShut {NoStop}%
\bibitem [{\citenamefont {Carleo}\ and\ \citenamefont
  {Troyer}(2016)}]{Carleo2016}%
  \BibitemOpen
  \bibfield  {author} {\bibinfo {author} {\bibfnamefont {G.}~\bibnamefont
  {Carleo}}\ and\ \bibinfo {author} {\bibfnamefont {M.}~\bibnamefont
  {Troyer}},\ }\href {https://arxiv.org/abs/1606.02318} {\bibfield  {journal}
  {\bibinfo  {journal} {arXiv:1606.02318}\ } (\bibinfo {year}
  {2016})}\BibitemShut {NoStop}%
\bibitem [{\citenamefont {Broecker}\ \emph {et~al.}(2016)\citenamefont
  {Broecker}, \citenamefont {Carrasquilla}, \citenamefont {Melko},\ and\
  \citenamefont {Trebst}}]{Broecker2016}%
  \BibitemOpen
  \bibfield  {author} {\bibinfo {author} {\bibfnamefont {P.}~\bibnamefont
  {Broecker}}, \bibinfo {author} {\bibfnamefont {J.}~\bibnamefont
  {Carrasquilla}}, \bibinfo {author} {\bibfnamefont {R.~G.}\ \bibnamefont
  {Melko}}, \ and\ \bibinfo {author} {\bibfnamefont {S.}~\bibnamefont
  {Trebst}},\ }\href {https://arxiv.org/abs/1608.07848} {\bibfield  {journal}
  {\bibinfo  {journal} {arXiv:1608.07848}\ } (\bibinfo {year}
  {2016})}\BibitemShut {NoStop}%
\bibitem [{\citenamefont {Chng}\ \emph {et~al.}(2016)\citenamefont {Chng},
  \citenamefont {Carrasquilla}, \citenamefont {Melko},\ and\ \citenamefont
  {Khatami}}]{Chng2016}%
  \BibitemOpen
  \bibfield  {author} {\bibinfo {author} {\bibfnamefont {K.}~\bibnamefont
  {Chng}}, \bibinfo {author} {\bibfnamefont {J.}~\bibnamefont {Carrasquilla}},
  \bibinfo {author} {\bibfnamefont {R.~G.}\ \bibnamefont {Melko}}, \ and\
  \bibinfo {author} {\bibfnamefont {E.}~\bibnamefont {Khatami}},\ }\href
  {http://arxiv.org/abs/1609.02552} {\bibfield  {journal} {\bibinfo  {journal}
  {arXiv:1609.02552}\ } (\bibinfo {year} {2016})}\BibitemShut {NoStop}%
\bibitem [{\citenamefont {Changlani}\ \emph {et~al.}(2015)\citenamefont
  {Changlani}, \citenamefont {Zheng},\ and\ \citenamefont
  {Wagner}}]{changlani2015}%
  \BibitemOpen
  \bibfield  {author} {\bibinfo {author} {\bibfnamefont {H.~J.}\ \bibnamefont
  {Changlani}}, \bibinfo {author} {\bibfnamefont {H.}~\bibnamefont {Zheng}}, \
  and\ \bibinfo {author} {\bibfnamefont {L.~K.}\ \bibnamefont {Wagner}},\
  }\href@noop {} {\bibfield  {journal} {\bibinfo  {journal} {The Journal of
  chemical physics}\ }\textbf {\bibinfo {volume} {143}},\ \bibinfo {pages}
  {102814} (\bibinfo {year} {2015})}\BibitemShut {NoStop}%
\bibitem [{\citenamefont {Barkema}\ and\ \citenamefont
  {Marko}(1993)}]{Barkema1993}%
  \BibitemOpen
  \bibfield  {author} {\bibinfo {author} {\bibfnamefont {G.~T.}\ \bibnamefont
  {Barkema}}\ and\ \bibinfo {author} {\bibfnamefont {J.~F.}\ \bibnamefont
  {Marko}},\ }\href {\doibase 10.1103/PhysRevLett.71.2070} {\bibfield
  {journal} {\bibinfo  {journal} {Phys. Rev. Lett.}\ }\textbf {\bibinfo
  {volume} {71}},\ \bibinfo {pages} {2070} (\bibinfo {year}
  {1993})}\BibitemShut {NoStop}%
\bibitem [{\citenamefont {Assaad}\ and\ \citenamefont
  {Evertz}(2008)}]{AssaadEvertz2008}%
  \BibitemOpen
  \bibfield  {author} {\bibinfo {author} {\bibfnamefont {F.}~\bibnamefont
  {Assaad}}\ and\ \bibinfo {author} {\bibfnamefont {H.}~\bibnamefont
  {Evertz}},\ }in\ \href {\doibase 10.1007/978-3-540-74686-7_10} {\emph
  {\bibinfo {booktitle} {Computational Many-Particle Physics}}},\ \bibinfo
  {series} {Lecture Notes in Physics}, Vol.\ \bibinfo {volume} {739},\ \bibinfo
  {editor} {edited by\ \bibinfo {editor} {\bibfnamefont {H.}~\bibnamefont
  {Fehske}}, \bibinfo {editor} {\bibfnamefont {R.}~\bibnamefont {Schneider}}, \
  and\ \bibinfo {editor} {\bibfnamefont {A.}~\bibnamefont {Wei{\ss}e}}}\
  (\bibinfo  {publisher} {Springer Berlin Heidelberg},\ \bibinfo {year}
  {2008})\ pp.\ \bibinfo {pages} {277--356}\BibitemShut {NoStop}%
\bibitem [{\citenamefont {Liu}\ \emph {et~al.}(2016)\citenamefont {Liu},
  \citenamefont {Shen}, \citenamefont {Qi}, \citenamefont {Meng},\ and\
  \citenamefont {Fu}}]{liu2016fermion}%
  \BibitemOpen
  \bibfield  {author} {\bibinfo {author} {\bibfnamefont {J.}~\bibnamefont
  {Liu}}, \bibinfo {author} {\bibfnamefont {H.}~\bibnamefont {Shen}}, \bibinfo
  {author} {\bibfnamefont {Y.}~\bibnamefont {Qi}}, \bibinfo {author}
  {\bibfnamefont {Z.~Y.}\ \bibnamefont {Meng}}, \ and\ \bibinfo {author}
  {\bibfnamefont {L.}~\bibnamefont {Fu}},\ }\href
  {https://arxiv.org/abs/1611.09364} {\bibfield  {journal} {\bibinfo  {journal}
  {arXiv:1611.09364}\ } (\bibinfo {year} {2016})}\BibitemShut {NoStop}%
\bibitem [{\citenamefont {Xu}\ \emph {et~al.}(2016)\citenamefont {Xu},
  \citenamefont {Qi}, \citenamefont {Liu}, \citenamefont {Fu},\ and\
  \citenamefont {Meng}}]{xu2016}%
  \BibitemOpen
  \bibfield  {author} {\bibinfo {author} {\bibfnamefont {X.-Y.}\ \bibnamefont
  {Xu}}, \bibinfo {author} {\bibfnamefont {Y.}~\bibnamefont {Qi}}, \bibinfo
  {author} {\bibfnamefont {J.}~\bibnamefont {Liu}}, \bibinfo {author}
  {\bibfnamefont {L.}~\bibnamefont {Fu}}, \ and\ \bibinfo {author}
  {\bibfnamefont {Z.-Y.}\ \bibnamefont {Meng}},\ }\href
  {https://arxiv.org/abs/1612.03804} {\bibfield  {journal} {\bibinfo  {journal}
  {arXiv:1612.03804}\ } (\bibinfo {year} {2016})}\BibitemShut {NoStop}%
\bibitem [{\citenamefont {Huang}\ and\ \citenamefont
  {Wang}(2016)}]{huang2016accelerate}%
  \BibitemOpen
  \bibfield  {author} {\bibinfo {author} {\bibfnamefont {L.}~\bibnamefont
  {Huang}}\ and\ \bibinfo {author} {\bibfnamefont {L.}~\bibnamefont {Wang}},\
  }\href@noop {} {\bibfield  {journal} {\bibinfo  {journal} {arXiv:1610.02746}\
  } (\bibinfo {year} {2016})}\BibitemShut {NoStop}%
\end{thebibliography}%

\end{document}